\documentclass[11pt,a4paper]{article}
\usepackage[english]{babel}
\usepackage[utf8]{inputenc}
\usepackage[T1]{fontenc}
\usepackage[a4paper, top=1in, bottom=1in, left=1.25in, right=1.25in]{geometry}

\usepackage[table,xcdraw]{xcolor}

\usepackage{verbatim}
\usepackage{multirow}
\usepackage{array, tabularx}
\usepackage{comment}
\usepackage{xr}
\usepackage{siunitx}
\usepackage{booktabs}
\usepackage{xspace}
\usepackage{float}
\usepackage{alltt}
\usepackage{makecell}
\usepackage{multicol}
\usepackage[super]{nth}
\usepackage{csquotes}

\usepackage[tagged, highstructure]{accessibility}
\usepackage{graphicx}
\usepackage{subcaption}
\usepackage{authblk}
\usepackage{amsmath,amssymb,amsthm}
\usepackage{url}
\usepackage{pifont}

\let\origref\ref
\def\ref#1{\textnormal{\origref{#1}}}

\title{Human/AI Collective Intelligence for Deliberative Democracy: A Human-Centred Design Approach }

\author[1]{Anna De Liddo*, https://orcid.org/0000-0003-0301-1154}
\author[1]{Lucas Anastasiou, https://orcid.org/0000-0002-1587-5104}
\author[2]{Simon Buckingham Shum$^*$, https://orcid.org/0000-0002-6334-7429}

\affil[1]{Knowledge Media Institute, The Open University, Milton Keynes, UK,\ \ E-mails:\{lucas.anastasiou, anna.deliddo\}@open.ac.uk}

\affil[2]{Connected Intelligence Centre, University of Technology Sydney, Sydney, Australia\\ E-mail: Simon.BuckinghamShum@uts.edu.au}

\date{\today}
\begin{document}

\maketitle

\begin{abstract}
This chapter introduces the concept of Collective Intelligence for Deliberative Democracy (CI4DD). We propose that the use of computational tools, specifically artificial intelligence to advance deliberative democracy, is an instantiation of a broader class of human-computer system designed to augment collective intelligence. Further, we argue for a fundamentally human-centred design approach to orchestrate how stakeholders can contribute meaningfully to shaping the artifacts and processes needed to create trustworthy DD processes. We first contextualise the key concepts of CI and the role of AI within it. We then detail our co-design methodology for identifying key challenges, refining user scenarios, and deriving technical implications. Two exemplar cases illustrate how user requirements from civic organisations were implemented with AI support and piloted in authentic contexts.
		\footnotetext[1]{$^*$ Corresponding author: Anna De Liddo, Knowledge Media Institute, The Open University, Milton Keynes, UK, E-mail: anna.deliddo@open.ac.uk}

{\bf Keywords}: Deliberative Democracy, Collective Intelligence, Human-AI Collaboration, Human-Centred Design, Civic Technology
\end{abstract}

\section{Introduction}

As we write this chapter in the final quarter of 2025, this tumultuous year has made all of us acutely aware of the fragility of democratic systems and the international rule of law. When we add to this the relentless breach of our planetary boundaries~\cite{richardson2023boundaries}, biodiversity collapse, the continued after-effects of the Covid-19 pandemic, and numerous other mutually interacting, degrading global systems, we do indeed appear to be wrestling with a ``global polycrisis'', defined by Lawrence,~\textit{et al.}~\cite{lawrence2024polycrisis} as, \textit{``the causal entanglement of crises in multiple global systems in ways that significantly degrade humanity’s prospects''}. 

Inserting itself into all of these systems is the explosive arrival of generative artificial intelligence, adding additional stressors: aggravating disinformation~\cite{dipto2024misinfo}, disrupting education~\cite{jensen2024genai} and employment~\cite{gmyrek2025ilo}, sparking debate on intellectual property law \cite{chesterman2025ip},  natural resource impacts \cite{ren2024llmenv,barker2025llmenv}, and precarious labour conditions for model training \cite{gray2019ghost}. Fully recognising that, like all technology, AI confronts us with ethical compromises and dilemmas, we remain equally intrigued by the extraordinary opportunities it opens up to ``augment human intellect'', to use the prescient 1960s language of Douglas Engelbart \cite{engelbart1962augment}. In a time when democracy is challenged by the polycrisis, we are curious to understand and harness the affordances of AI, just as it holds significant potential to augment strategic planning \cite{doshi2025aistrategy} and learning amidst the polycrisis \cite{buckinghamshum2025flourish}.  

Consequently, in this chapter we advocate for the responsible use of AI to advance deliberative democracy (DD). Furthermore, we see this as an instance of the broader class of human-computer systems designed to augment collective intelligence (CI), with hybrid forms of human/AI CI now emerging. If this holds, it follows that DD research and practice can both learn from CI, and contribute back to it. Further, we argue for a fundamentally human-centred design (HCD) approach to orchestrate how stakeholders can contribute meaningfully to shaping the artifacts and processes needed to create a trustworthy DD process. To achieve this, the paper presents a structured co-design process for developing AI tools that enhance democratic deliberation processes. 

In the context of a European project involving diverse DD organisations, we detail a collaborative scenario development process through four phases — \textit{Contextualization, Community Challenges, Scenario Co-creation} and \textit{Validation}. From this we identified key ``Points of Struggle'' in existing deliberative practices, and translated user aspirations into technically feasible system requirements. Our methodology bridges the gap between theoretical AI capabilities and practical deliberative needs, adhering to the principle of augmenting rather than replacing human capabilities. We demonstrate how this approach leads to contextually relevant, user-centric AI solutions that address real-world challenges in citizen participation, collective understanding, transparency, and scalability of deliberative processes. 

The chapter first  contextualises the key concepts of CI and the role of AI within it (Sec.~\ref{sec:CI_for_DD}).
We then detail our HCD methodology and its outcomes, including key challenges and a catalogue of system requirements (Sec.~\ref{sec:methodology}). Subsequently, we present two exemplar cases of how these requirements can be implemented: BCause (Sec.~\ref{sec:Bcause}) and DemocraticReflection (Sec.~\ref{sec:dem_reflection}). We conclude by summarising how these exemplars address the challenge of designing hybrid collective intelligence for deliberative democracy (Sec.~\ref{sec:conclusions}).

\section{CI, AI and DD}
\label{sec:CI_for_DD}
Fields as diverse and intersecting as organisation science, cognitive science, computer science and neuroscience are converging on the importance of Collective Intelligence (CI), ranging in scale from small teams to companies, and to global networks~\cite{malone2015handbook}. In the editorial to the inaugural edition of CI Journal, Flack et al.~\cite{Flack2022} introduce CI as follows:

\begin{displayquote}
We can find collective intelligence in any system in which entities collectively, but not necessarily cooperatively, act in ways that seem intelligent. Often—but not always—the group’s intelligence is greater than the intelligence of individual entities in the collective.
\end{displayquote}

Central to the concept of CI is the premise that intelligence cannot be restricted to what happens in an individual brain, but rather, intelligence (including cognition and memory) is distributed socially across agents and materially across artifacts, both physical and computational~\cite{Hollan2000,ONeill2023}. Consequently, CI considers more than the collective ability of people’s minds, with online platforms making new forms of discourse and coordination possible~\cite{Gupta2023,DeLiddo2011,Suran2020,vanGelder2020}. AI adds machine actors to the network, with human-agent teaming research clarifying the conditions under which people come to trust AI agents as members of the team~\cite{ONeill2020,Seeber2020}.
The explosive arrival of large language models (LLMs), particularly their integration with a conversational user interface (marked most saliently by ChatGPT’s launch in Nov. 2022) is the most recent advance, opening new possibilities for human-computer creativity~\cite{Heyman2024}, CI more broadly~\cite{Burton2024} and ``extended minds''~\cite{Clark2025}.

Whilst there are distinctive features of DD that do not necessarily hold in other contexts (e.g., the particular kinds of stakeholders, or the common need for policy outcomes), we propose that CI concepts are helpful in framing the challenge of convening trustworthy, computer-supported, DD processes. For instance, in a detailed survey of CI online platforms, Suran et al.~\cite{Suran2020} taxonomise approaches in terms of the \textit{individuals} they bring together (including their motivation, diversity, critical mass), through \textit{coordination/collaboration} activities (including the scope for emergence, trust, task allocation), enabled by different forms of \textit{online communication} (including user roles, knowledge aggregation, decentralisation). Their survey includes a range of democracy-enhancing, citizen-centric cases within this framework~\cite{Furtado2010,Iandoli2009,Kosmidis2018}.

Our previous research has proposed that complex sociotechnical problems need a specific type of discourse-centric CI, which we termed \textit{Contested Collective Intelligence (CCI)}~\cite{DeLiddo2011}. Divergence in opinion is intrinsic to the predicament, and computational support assists in making sense of this. The nature of the CCI design challenge has been characterised as a spectrum of ``pain points'' that may be addressed by different forms of deliberation analytics ~\cite{shum2014dcla}. CCI deliberation platforms have been piloted in contexts including education, civic engagement, and strategic/spatial planning. 

\begin{figure}
    \centering
    \caption{Collective Intelligence for Deliberative Democracy Research (CI4DD)}
    \includegraphics[width=0.5\linewidth]{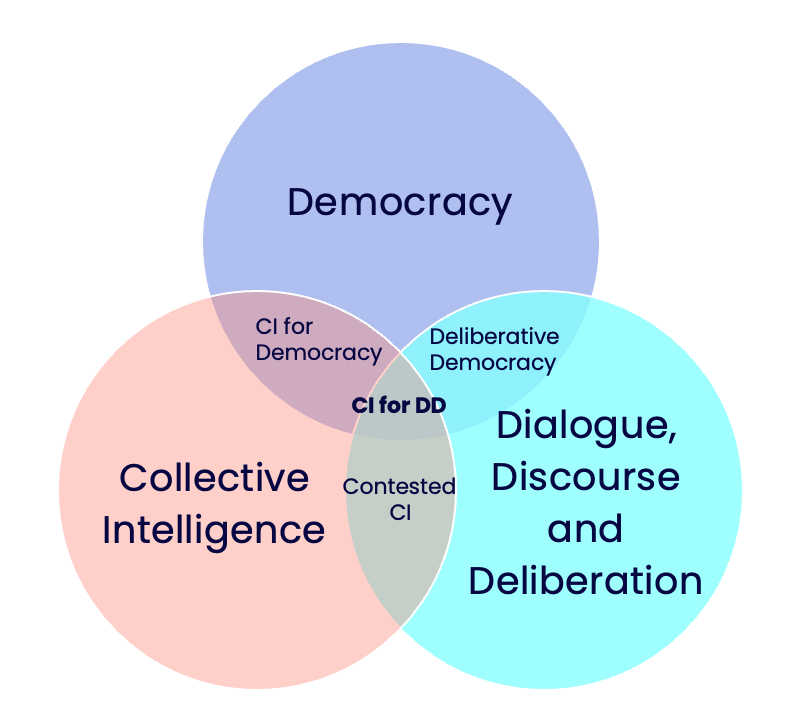}    
    \label{CI4DD}
\end{figure}

Similarly, in the realm of CI systems for democracy, a subset of systems exists that, by structuring dialogue and deliberation, produces new forms of collective intelligence, which lead to democratic outcomes. We term this CI for Deliberative Democracy (CI4DD), a crucial research endeavor to advance digitally mediated democracy, which we locate at the intersection of Democracy, Deliberation and Collective intelligence research (Figure~\ref{CI4DD}). 

Gupta et al.~\cite{Gupta2023} ask: \textit{``How do we know that such a sociotechnical system as a whole, consisting of a complex web of hundreds of human–machine interactions, is exhibiting CI?''} They argue for sociotechnical architectures capable of sustaining ``collective memory, attention, and reasoning.'' What does it mean to ``exhibit CI'' in the context of DD? So, we may contextualise this question, as follows: \textit{``How do we know that a DD sociotechnical system as a whole, consisting of a complex web of hundreds of human–machine interactions, is exhibiting democratic results, impacts and behaviours?''} In this chapter, we will illustrate how the above ‘cognitive faculties’ of a CI4DD platform can be instantiated in a trustworthy manner. This revolves around giving stakeholders a meaningful voice in the design process.

%
\subsection{A Comparative Analysis of Deliberative Technology Platforms}

Arguably the majority of DD work using online tools still works with generic tools for synchronous work (video conferencing plus online meeting tools) and asynchronous work (threaded discussions, collaborative documents, digital sticky-note boards, etc). An example is a series of online workshops using DD processes to scaffold a "deliberative mini-public" to draft AI ethics principles, where the key digital tools were Google Docs and Zoom \cite{swist2024}. Within the last 3 years, of course, the rise of LLMs sees AI increasingly present in such everyday tools, enabling automated textual summarisation, text/image generation, sticky-note clustering, thematic analysis, and chatbots grounded in a DD document corpus. These are all legitimately examples of ``AI for DD''.

However, our focus in this chapter is on the use of AI in platforms with features specifically tuned to support DD processes and work practices, with particular interest in building ``CI''. To situate our work within the broader CI4DD landscape, Table \ref{tab:comparison} provides a comparison across prominent platforms. We selected these platforms as exemplars of different deliberative modalities: dialectical approaches that structure arguments explicitly (Kialo\footnote{https://www.kialo.com/}, BCause\footnote{https://bcause.app/}), aggregative systems that synthesize distributed input (Pol.is\footnote{https://pol.is/}, DemocraticReflection\footnote{https://democratic-reflection.web.app/en}), consensus-oriented tools for group decision-making (Loomio\footnote{https://www.loomio.com/}), and municipal governance platforms (Adhocracy+\footnote{https://adhocracy.plus/}, Consul\footnote{https://consuldemocracy.org/}). This comparison uses the analytical dimensions discussed above to highlight how different sociotechnical systems attempt to resolve the fundamental challenges of achieving effective, inclusive, and scalable deliberative democracy.

\begin{table}[ht]
\centering
\caption{A Comparative Analysis of Deliberative Technology Platforms}
\footnotesize
\begin{tabular}{>{\raggedright\arraybackslash}p{1.4cm}p{1.5cm}p{1.5cm}p{1.5cm}p{1.5cm}p{1.5cm}p{1.5cm}p{1.5cm}}
\toprule
\textbf{Platform} & \textbf{BCause} & \textbf{Democratic\-Reflection} & \textbf{Kialo} & \textbf{Pol.is} & \textbf{Loomio} & \textbf{Adhocracy+} & \textbf{Consul} \\
\midrule
Core Deliberation Mode & Dialectical (Argument Map) & Aggregative (Real-time) & Dialectical (Argument Map) & Aggregative (Clustering) & Consensus (Decision) & Modular (Multi-modal) & Modular (Multi-modal) \\
\midrule
AI role &  Structuring \& sensemaking & Sensemaking \& facilitation & None & ML clustering & None & None & None (planned) \\
\midrule
Key features & Transcript import, arg. network, theme maps & Reflection cards, dashboards, AI questions & Pro/con tree, sunburst viz & Statement voting, group viz & Threaded discuss., polls, decisions & Idea submit, budgeting, geo-map & Proposals, debates, voting, budgeting \\
\midrule
Primary Use Case & Civic engagement, policy & Live events, consultations & Education, debate & Policy-making, large-scale & Governance, collectives & Municipal gov., civic & Municipal gov., institutions \\
\bottomrule
\end{tabular}%
\label{tab:comparison}
\end{table}

This comparative analysis reveals distinct gaps in the CI4DD technology landscape that our exemplar systems are designed to address. While many platforms excel at either highly structured asynchronous debate (e.g., Kialo) or large-scale asynchronous opinion aggregation (e.g., Pol.is), there is a notable lack of technologies designed to bridge different deliberative modalities — synchronous and asynchronous, online and offline. \textit{BCause} targets this specific gap by addressing the \textit{Integration} struggle. Its novel contribution lies in using AI to transform an unstructured, synchronous conversation (a meeting transcript) into a structured, asynchronous deliberative artifact (an argument map). This creates an analysable memorable data point from an ephemeral event, connecting offline deliberation with an ongoing online process in a way not prominent in the other platforms reviewed. \textit{DemocraticReflection} addresses a different, yet equally critical, challenge: fostering \textit{Collective Understanding} and \textit{Inclusivity} during a live, synchronous event. While most platforms are designed for asynchronous use or post-event analysis, DemocraticReflection provides a real-time feedback loop. It allows audience reflections to be captured, analysed, and fed back into the expert discourse as it happens, augmented by AI-as-facilitator. This demonstrates how our human-centered design approach led to the development of specific, context-aware solutions that address nuanced ``Points of Struggle'' (see Section~\ref{sec:results_pos}) often overlooked by more general-purpose platforms.


\section{A CI4DD co-design process}
\label{sec:methodology}

To elicit user requirements and co-create digital solutions for AI-augmented deliberative democracy, we employed a co-design methodology that emphasises participatory engagement with end-user communities. This approach was designed to bridge the gap between technological capabilities and the practical needs of democratic practitioners, ensuring that the resulting solutions would be both technically feasible and contextually relevant. The Horizon Europe’s ORBIS project \footnote{https://orbis-project.eu/} on ``Augmenting participation, co-creation, trust and transparency in Deliberative Democracy at all Scales'' provided the necessary context and human engagement for the co-design. ORBIS is one of five international projects funded as part of the \textit{AI, Big Data and Democracy} European Task Force to advance the understanding, development and real life impact of AI innovations for democracy. 

Our work is in partnership with democratic innovation initiatives and organisations including \textit{The Centre for European Policy Studies in Bruxelles (CEPS)}\footnote{https://www.ceps.eu/}, a leading think tank and forum for debate on EU affairs to inform policy making; \textit{The Democracy and Culture Foundation (DCF)}\footnote{https://www.democracyculturefoundation.org/} an organisation formed to empower society through citizen engagement and better governance; and \textit{Re-Imagine Europa (RIE)}\footnote{https://re-imagine.eu/}, a non-partisan think-tank currently focused on a European initiative called \textit{Future4Citizens} (\#F4C) to foster deliberation and democratic participation in Europe through a community-based and narrative approach. These organisations provide access to both problems and communities for developing and evaluating the CI4DD co-design process. 

\begin{figure}
\centering
\caption{The three-phase co-design methodology: (Community Challenges, Scenarios Co-creation, and Validation) showing key stakeholders (Who), Objectives, and Outputs for each phase in the development of AI-augmented deliberative democracy tools.}
\includegraphics[width=1.1\textwidth]{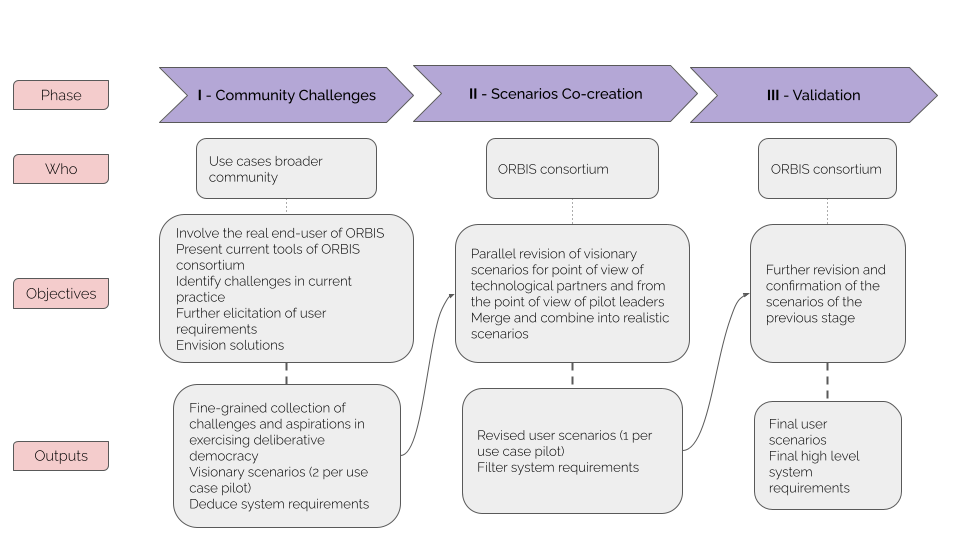}
\label{fig:four_stages_co_design}
\end{figure}

Our methodology unfolded in three phases, as summarised in Figure~\ref{fig:four_stages_co_design}.

\subsection{Phase I: Eliciting Community Needs and Aspirations}

To ground our research in real-world needs, we initiated our co-design process by engaging directly with the intended end-users of deliberative democracy systems. We conducted five workshops with 35 participants from key stakeholder groups, including NGOs, advocacy groups, and civic society organisations involved in citizen-led policymaking. The goal was to understand their challenges and aspirations for technological support in deliberative processes. Each two-hour virtual workshop was a collaborative session using tools like Miro~\footnote{https://miro.com/} and Mentimeter~\footnote{https://www.mentimeter.com/} to capture ideas and feedback. This process generated over 10 hours of discussions, alongside numerous digital artifacts like virtual sticky notes and poll results. To analyze this rich dataset, we employed a hybrid methodology, combining a deductive thematic analysis with a bottom-up, grounded theory approach~\cite{glaser2017discovery,chen2022theory,Braun2006}. This involved transcribing the discussions and systematically coding the data from all sources to identify recurring patterns and themes. This analysis distilled the communities' high-level desires and pain points, which directly informed the next phase of co-creating user scenarios and, ultimately, a catalogue of system requirements for an AI-enhanced Deliberative Democracy platform.

\subsection{Results: Key points of struggle (PoS)}\label{sec:results_pos}


The open coding analysis was conducted through an iterative process involving researchers reviewing workshop transcripts, (virtual) sticky notes, poll results and other notes. Initial codes were generated inductively focusing on specific challenges mentioned by participants. Subsequent rounds aimed to group codes into higher-order themes, capture emerging patterns, and continuously refine categories until data saturation was reached and no new significant themes emerged. The result was the identification of four key points of struggle that permeated across the different deliberative phases (thematic lenses of observation). These points proved to be \textit{robust, recurring elements} across the spectrum of democratic innovation organisations focused on advocacy and citizen-led policy making (so should not be considered an exhaustive list covering all points). 

\begin{itemize}
    \item \textbf{Achieving True Representation and Inclusivity}: Despite the recognition of the importance of ensuring that all voices are represented in deliberative processes, participants highlighted the challenges of overcoming under-representation and ensuring that all perspectives are heard. This challenge is also reported extensively by literature, for example in Karpowitz and Raphael~\cite{Karpowitz2016}, and Bochel et al~\cite{Bochel2008}. The challenge of true representation and inclusivity includes actively engaging with (and incorporating the opinions of individuals) and communities that may be traditionally marginalised or have limited access to participation opportunities. A specific need to investigate issues of intersectionalities was also pointed out. 
    
    \item \textbf{Collective Understanding and Shared Reality}: The concept of ``collective sensemaking''~\cite{de2010capturing}, ``shared reality''~\cite{Dugas2018}, and ``social representation''~\cite{Breakwell2014} emerged as a crucial aspect of deliberative processes. Creating a common ground where participants can understand each other's viewpoints and feel represented is essential for fostering meaningful dialogue and achieving informed decision-making. This shared understanding of knowledge and experiences is crucial for bridging social divides and promoting inclusive governance~\cite{Dugas2018}.
    
    \item \textbf{Clarity and Transparency in Process and Outcome}: Transparent explanation and a clear understanding of the flow of results are essential for ensuring the legitimacy and effectiveness of deliberative processes~\cite{greene2000evanescent}. This includes providing clear explanations of decision-making processes (as highlighted for example in~\cite{Petts2001}, showcasing the rationale behind outcomes, and demonstrating the impact of deliberations on policy decisions. Policymakers must be able to trust the outcomes of deliberative processes and understand how certain decisions were reached; which provides legitimacy to the decisions made~\cite{Parkinson2003}.
    
    \item \textbf{Integration and Scalability}: As emphasised by Klein~\cite{Klein2012}, the ability to integrate deliberative processes with existing tools and platforms is crucial for their long-term sustainability and scalability. This includes supporting multiple languages, as the EU is a multilingual organisation with 24 official languages, and enabling seamless integration with various stakeholders and decision-making bodies. The ability to scale deliberative processes horizontally (across different communities), vertically (to address complex issues), and in-depth (to incorporate more detailed information) is essential for addressing a wider range of challenges~\cite{delBook2012} and achieving greater impact~\cite{Shortall2022}.
    
\end{itemize}

As common denominators across the diverse contexts of consulted communities, these four `Points of Struggle' highlight fundamental challenges if DD platforms are to be effective in real-world applications.

\subsection{Phase II: User Scenario Co-Creation}

Building on insights from earlier phases, we developed twelve ``seed user scenarios'' intended to provoke thought and encourage ambitious thinking. These initial scenarios, exceed the ORBIS project scope, and served as valuable starting points for a co-creation process aimed at refining them into realistic and actionable use cases. Each scenario was crafted using a structured format that included detailed personas and a six-part narrative structure, supplemented by annotations referencing a predefined set of deliberative functionalities.

After the scenarios were defined two parallel workshops were convened to evaluate and revise them. We ran a technology focused workshop, involving deliberaive technology experts to conduct a structured review of all twelve scenarios, annotating them for necessity, desirability, and technical feasibility. After this the scenarios were amended and reworked into technically viable versions, also identifying system-level requirements that could realistically be implemented.

Simultaneously, a user community workshop was conducted involving end-user representatives, who reviewed the same scenarios from a usability and relevance perspective. Through facilitated discussions and collaborative evaluation, each use case team combined two scenarios into a single preferred version, incorporating feedback from peers and other stakeholders.

A subsequent plenary session brought both groups together to reconcile technical feasibility with user needs. Participants presented revised scenarios and engaged in moderated discussions to identify common ground and resolve differences. This co-creative negotiation resulted in compromised yet balanced final versions of the scenarios, ensuring that each retained operational relevance while remaining within technological constraints. Ultimately, the process transformed the original twelve visionary concepts into six realistic and implementable scenarios that have underpinned the ORBIS project’s technical design and development. In the following section, we will exemplify results of the development and application of such scenarios with two deliberation technologies applied in real-world contexts.

\subsection{Phase III: Use Case Scenarios Validation}

\begin{figure}
    \centering
    \caption{``Engaging teenagers'' DD user scenario co-created by civic organisations. The scenario follows a structured narrative format: (top panel) begins with \textit{Persona} description and \textit{Context} setting, establishing the stakeholder and situation; (middle panel) progresses through \textit{Challenge} identification and \textit{Objective} definition, clarifying what needs to be achieved; (bottom panel) details the \textit{Actions} taken and \textit{Outcomes} realized, showing how the deliberative technology enables the solution. Coloured annotations on the right link specific narrative elements to system requirements (e.g., UIE.2, DAV.1) detailed in the Appendix, creating traceability between user needs and technical specifications.}
    \includegraphics[width=0.95\linewidth]{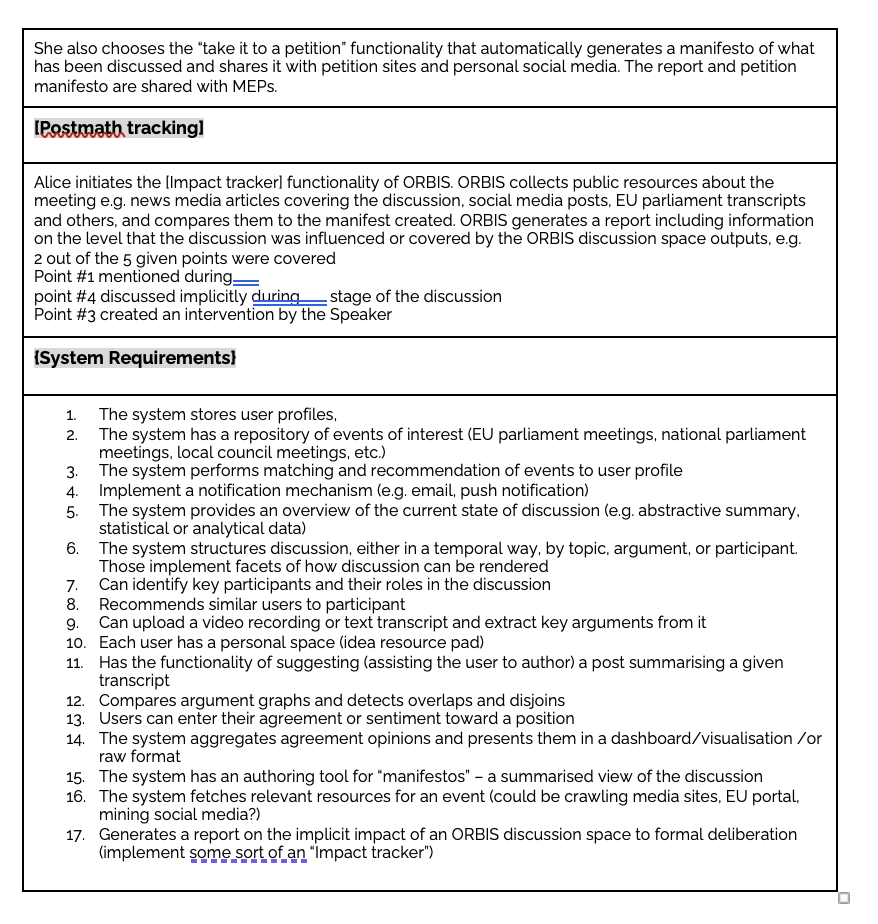}    
    \label{fig:Example_Teens_Scenario}
\end{figure}

The final phase involves distributing the scenarios to all stakeholders for review and validation. Use case leaders conducted internal consultations within their organisations to verify alignment with institutional needs, while technology partners provided final assessments of technical feasibility. This iterative validation process, though primarily qualitative due to participant numbers, was intrinsically embedded within the co-design framework, ensuring rigorous collaborative assessment and consensus regarding feasibility and authenticity. The co-created user scenarios offered descriptions of user journeys and interactions with fictitious deliberation systems. While some of them may not yet be technologically feasible, they encompassed the full set of functionalities expected in a mature deliberative solution as envisioned by participants. 
Figure \ref{fig:Example_Teens_Scenario} shows one of these user scenarios, related to engaging teenagers in DD processes. 

\begin{figure}
    \centering
    \includegraphics[width=0.95\linewidth]{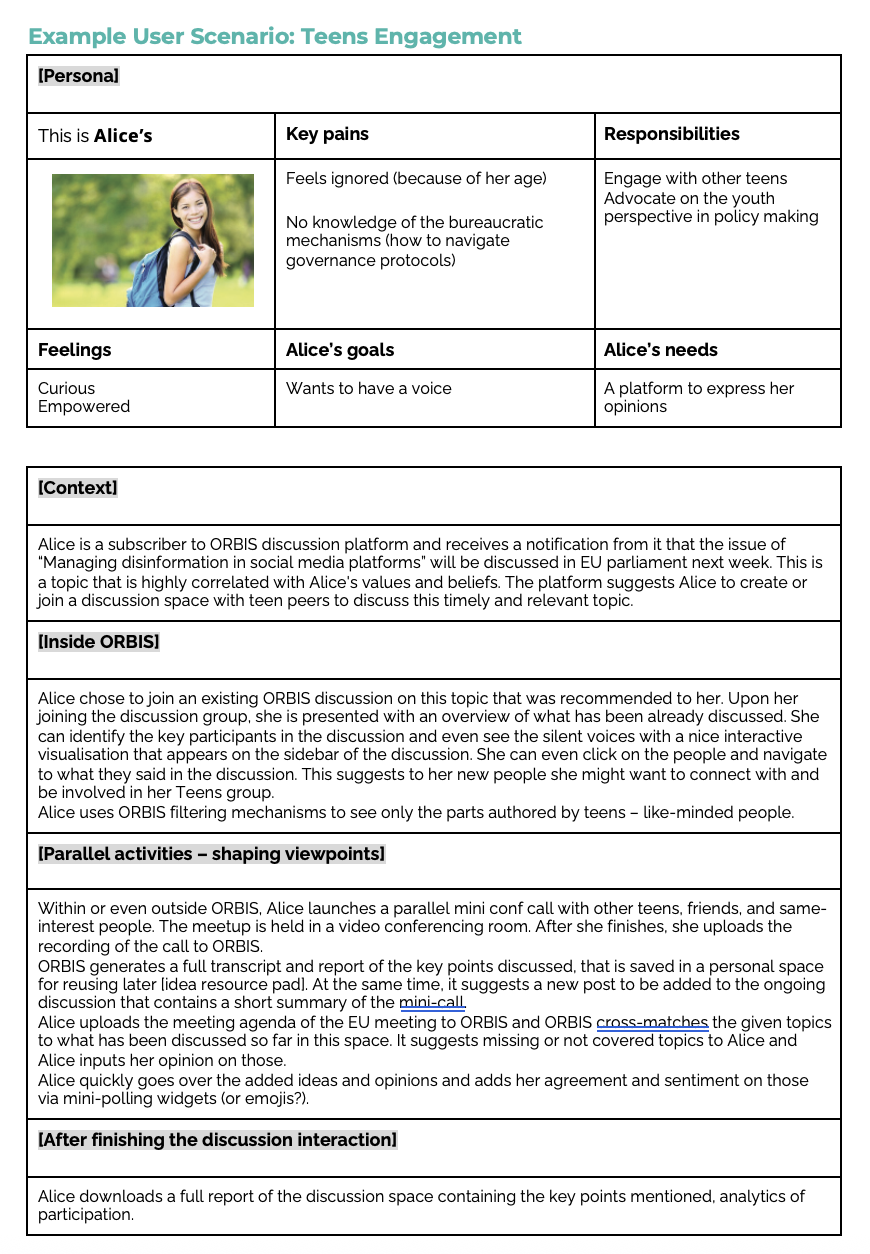}
\end{figure}


\subsubsection{Higher-Level System Requirements}

It is one thing to envision user scenarios, but these must at some point be translated into technical reality. The scenarios are used to elicit a set of high-level system requirements. 
At the bottom of the example user scenario (Figure~\ref{fig:Example_Teens_Scenario}) the system requirements identified for that scenario are listed. From the 12 scenarios, 121 system requirements were derived. We initiated an open coding process, iteratively tagging each one of them, merging them according to their common functionalities, from which we distilled a subset of 36 requirements (see Appendix). These were then further distilled into six  categories which guided platform development.

\begin{enumerate}
    \item{\textbf{User Interaction and Engagement (UIE):}} This category covers the overall user experience, focusing on personalization and flexibility. Users expressed a need for systems that support user profiles, provide personal spaces for ideas, offer personalized recommendations, and include user-friendly visualizations for data exploration and real-time discussion support.
    \item{\textbf{Discussion Analysis and Visualization (DAV):}} This addresses the user's high expectations for data analysis and sensemaking. Key requirements include the ability to automatically cluster discussions into themes, analyze and compare arguments, identify key actors, and summarize discourse in real-time through interactive visualizations.
    \item{\textbf{Moderation and Assistance (MA):}} Focusing on support for expert facilitators, these requirements aim to enhance the moderation of deliberation. This includes tools for gathering feedback efficiently, creating structured records of discussions, detecting controversy, and providing automated indicators to assist moderators in guiding the process.
    \item{\textbf{Reports and Summarization (RS):}} These requirements center on generating evidence-based insights from deliberations. Users need tools that can produce statistical analyses and generate summary reports in various styles, while crucially maintaining clear links back to the original source data to ensure provenance.
    \item{\textbf{Collaborative Features (CF):}} To foster collective sensemaking, this category includes features that allow users to actively co-create meaning. This involves functionalities for collaborative filtering, grouping, prioritizing, and editing of deliberative data, as well as mechanisms for voting and connecting to external knowledge.
    \item{\textbf{Multi-Phase Deliberations (MPD):}} This reflects the understanding that deliberation is a process that unfolds across different times, spaces, and modalities. Requirements include support for tracking the process across various stages, aggregating results from surveys and polls, and generating policy proposals from discussion outcomes.
\end{enumerate}


To summarise, this codesign requirements elicitation process enabled citizens and deliberative democracy actors (NGOs, civic society organisations, governmental institutions etc) to identify their most acute `points of struggle', and articulate user stories that brought to life what CI4DD could mean for them. From these rich narratives we derived more abstract, higher-level requirements to guide CI4DD platform designers. As explained next, some of the stakeholders' envisioned capabilities require AI as part of the overall sociotechnical solution.


\section{Hybrid Deliberation via BCause: Bridging Structured Asynchronous Discussion with Unstructured Synchronous Conversations}
\label{sec:Bcause}

\subsection{BCause: Platform Description}

The BCause platform~\cite{Anastasiou2023}\footnote{https://bcause.app} exemplifies a human-centred approach to AI-augmented asynchronous deliberation. Developed by the IDea\footnote{https://idea.kmi.open.ac.uk/} group at the Open University, BCause is a structured online discussion system that supports distributed decision-making through argumentative dialogue enhanced by AI capabilities. The platform addresses the challenge of scaling deliberative democracy while maintaining the quality and structure necessary for meaningful civic engagement.

\begin{figure}
    \caption{Snapshot of BCause Discussion Interface: The focal \textit{Question} is at the top, and the left panel generates a succinct summary of the \textit{most contested and opposed positions}. These are listed in the central panel, each with related \textit{Arguments} organised into \textit{Cons} (left) and \textit{Pros} (right). The overview tree on the right supports orientation and navigation.}
    \centering
    \includegraphics[width=1\linewidth]{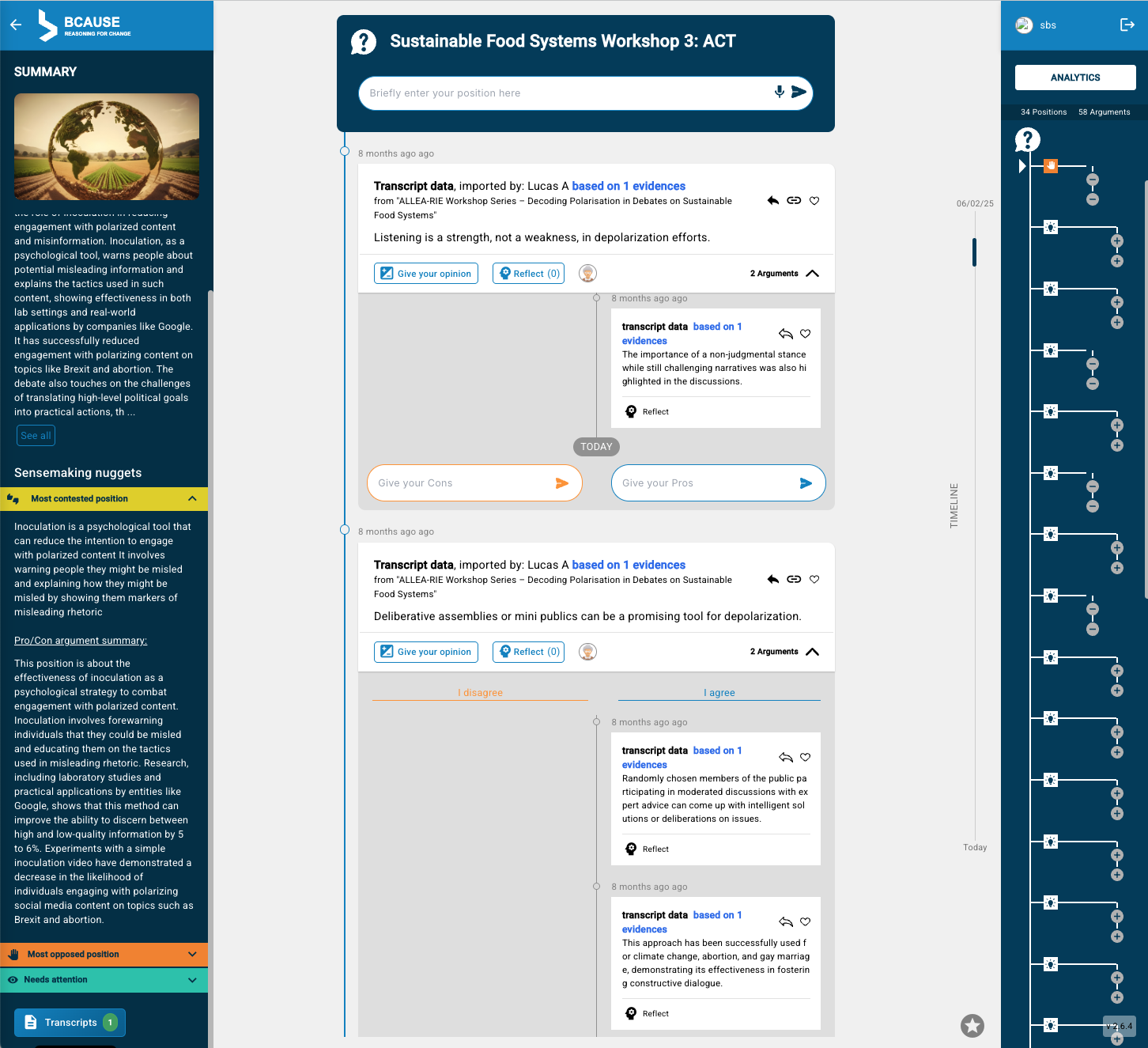}
    \label{bcause image}
\end{figure}

The core functionality centres on argumentative dialogue structure that helps participants engage in reasoned dialogue. For that, BCause employs a light IBIS (Issue-based Information System) model to structure participants' contributions around issues, positions and arguments (Figure \ref{bcause image}). 
The affordances of IBIS-based deliberation tools for improving the quality of deliberation are well documented; by explicitly structuring contributions into their argumentation role, the deliberation process becomes clearer, more rational and reveals the agreement and contention points of the group~\cite{Iandoli2009,DeLiddo2011}.
The platform's interface which separates pro and con arguments, is paired with a navigable argument tree on the right that provides a clear overview of the debate and summaries or other insights on the left. The platform has multiple engagement mechanisms: users can contribute detailed arguments, provide quick feedback through reflection mechanisms, or signify their agreement to given positions. At the same time, the system maintains clear authorship and argument provenance, ensuring transparency in how collective understanding emerges from individual contributions~\cite{anastasiou2023Thesis}.

The transition from theoretical requirements (section \ref{sec:methodology}) to practical implementation of BCause in the ORBIS project required careful consideration of how BCause could be extended to address the specific challenges identified through our co-design process. The RIE Europe4Citizens initiative presented particularly compelling contexts for testing BCause's capacity to bridge offline and online deliberative spaces while maintaining democratic legitimacy and fostering collective intelligence. RIE as an organisation exemplifies the core tension identified in our Points of Struggle analysis: how to achieve meaningful Collective Understanding and Shared Reality while ensuring Integration and Scalability without compromising True Representation and Clarity/Transparency. The Europe4Citizens dialogues, conducted across multiple European contexts, required a system capable of preserving the nuanced discussions occurring in face-to-face citizen dialogues while enabling broader online participation.

These requirements directly informed three critical extensions to the BCause platform: (1) AI-augmented transcript import functionality to preserve and structure offline deliberative content, (2) enhanced policy recommendation distillation capabilities, and (3) advanced argument clustering and visualization tools. 

\subsection{AI-augmented transcript import}

A distinctive feature of BCause is a human-AI collaboration process enabling the discussion curator to distill the key questions, ideas and arguments arising in face-to-face or virtual meetings, migrating these into online BCause discussions. The transcripts from a meeting (in-person or online, recorded with participant permission) are converted into BCause deliberation structures via a three-step human-in-the-loop workflow:
\begin{enumerate}
    \item \textbf{AI analysis of meeting transcript:} an Argument Mining pipeline uses a supervised machine learning approach (a fine-tuned DeBERTa transformer model) to automatically detect and categorise discussion elements into the IBIS schema (Issues, Positions and Arguments)
    \item \textbf{AI result visualisation and initial approval:} the discussion curator reviews and if necessary edits the IBIS structure
    \item \textbf{Merge with BCause for approval:} the curator confirms coherent threading of the new contributions into the existing BCause forum.
\end{enumerate}

This captures insights from naturalistic face-to-face/online meetings (i.e. relatively unstructured spoken transcripts), ensuring that the  AI structuring of the conversation has been executed transparently and accurately, thus enabling continuity of the discussion in the new BCause modality and slower, asynchronous tempo. 

\subsection{Policy Recommendations Distillation}

For the RIE Europe4Citizens use case, BCause was extended with enhanced policy synthesis capabilities. The platform now automatically identifies \textit{policy-relevant} argumentative components and generates structured recommendations that are connected with existing EU policy frameworks.

\begin{figure}
    \centering
    \caption{The BCause policy clusters dashboard, illustrating the main clusters from the `Sustainable Food Systems Workshop'. The interface organizes participants into distinct `Key Positions' with each one supported by `Supporting Claims' grounded in a `Originating Transcript Context'}
    \includegraphics[width=1\linewidth]{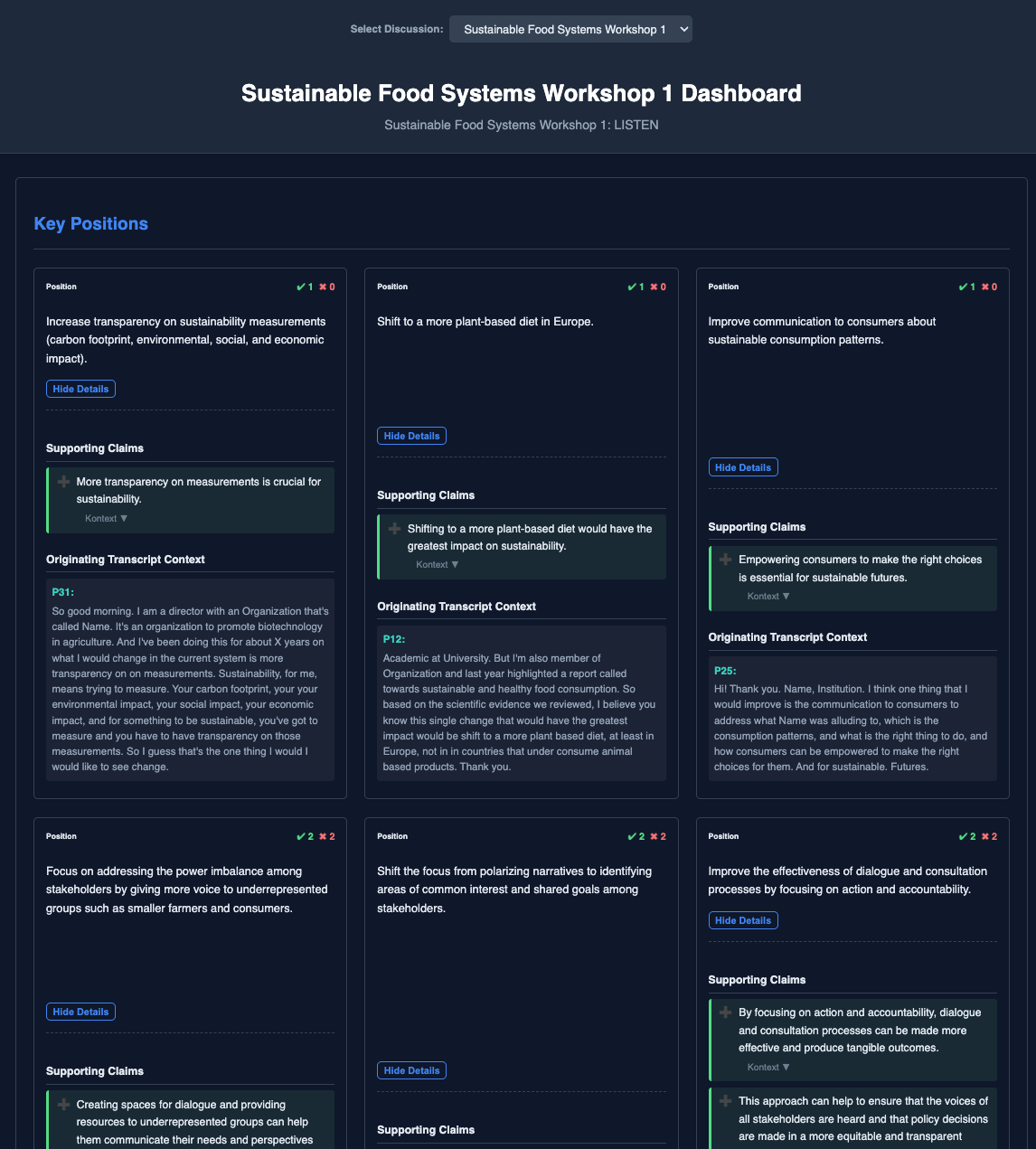}    
    \label{fig:BCause_Policy_dash_1}
\end{figure}

This is achieved through a multi-stage AI pipeline, whose key steps are:

\begin{enumerate}
    \item \textbf{Group similar arguments:} a Feedback Aggregator uses a Fuzzy C-means clustering algorithm to group semantically similar arguments
    \item \textbf{Label these clusters:} a Generative LLM assigns a title and short description to each cluster, and synthesises the key points in the cluster into a coherent policy recommendation
    \item \textbf{Enable policy analyst to explore results:} an interactive dashboard, e.g., Figure~\ref{fig:BCause_Policy_dash_1} has generated a set of recommendations (IBIS positions) with their supporting claims (IBIS Arguments) related to \textit{Sustainable Food Systems}, with links back to the originating transcript segments to ensure full provenance and transparency.
\end{enumerate}

This addresses the Integration and Scalability challenge by providing outputs that policymakers can directly incorporate into their deliberative processes. The capability of the policy recommendation feature to track consensus formations, identify areas of disagreement and generate nuanced policy recommendations, also proved particularly valuable for Young Thinkers events, where diverse perspectives needed to be synthesised into actionable input for EU decision makers.

\subsection{Advanced clustering and vizualisations}

New visualization tools were developed to represent the complexity of multi-stakeholder deliberations.

\subsubsection{Transcript Argument Markup and Argument network creation}

\begin{figure}
    \centering
     \caption{Identified argumentative components (claims and premises) are shown as markup in BCause transcript viewer.}
    \includegraphics[width=0.99\linewidth]{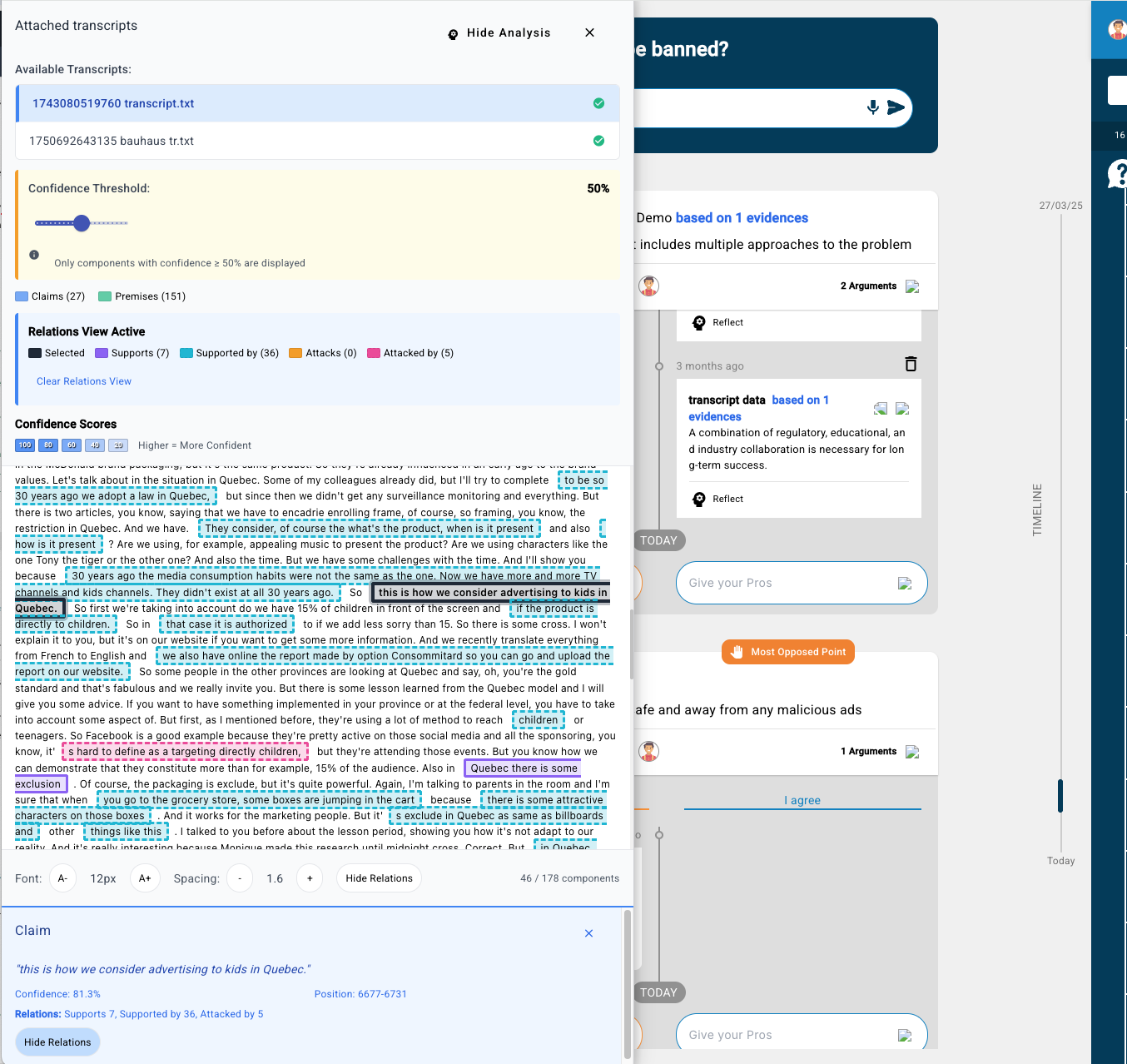}   
    \label{fig:transcript_markup}
\end{figure}

Uploaded transcripts undergo automated argument component identification, where the AI system marks up claims, evidence, and reasoning structures while preserving original authorship. As seen in Figure~\ref{fig:transcript_markup}\footnote{Example from https://bcause.app/discussions/-OMMdBg09OoAPTFhifkR}, the BCause transcript viewer does not just display text; but rather highlights argumentative components (e.g. claims, premises) identified by AI in distinct colours and their (argumentative) relations with other text snippets of the same transcript. This markup is the direct output of the Argument(ation) mining pipeline described earlier. This markup process aligns with the high-level system requirement for ``Discussion Analysis and Visualisation'' (DAV), specifically enabling transparent analysis of complex deliberative content. The resulting argument network combines contributions from both offline transcripts and online posts, creating a unified knowledge structure that addresses the Collective Understanding Point of Struggle by making implicit argumentative relationships explicit. This network tackles the \textit{Transparency} challenge by showing participants not just what the community thinks, but how that understanding emerged from individual contributions.

\subsubsection{Interactive clustering}
\begin{figure}[ht]
    \centering
    \caption{BCause interactive clustering analytics page with Voronoi (circle pack) mode selected variable number of clusters}
    \begin{subfigure}{0.3\textwidth}
        \centering
        \caption{2-Clusters}
        \includegraphics[width=\textwidth]{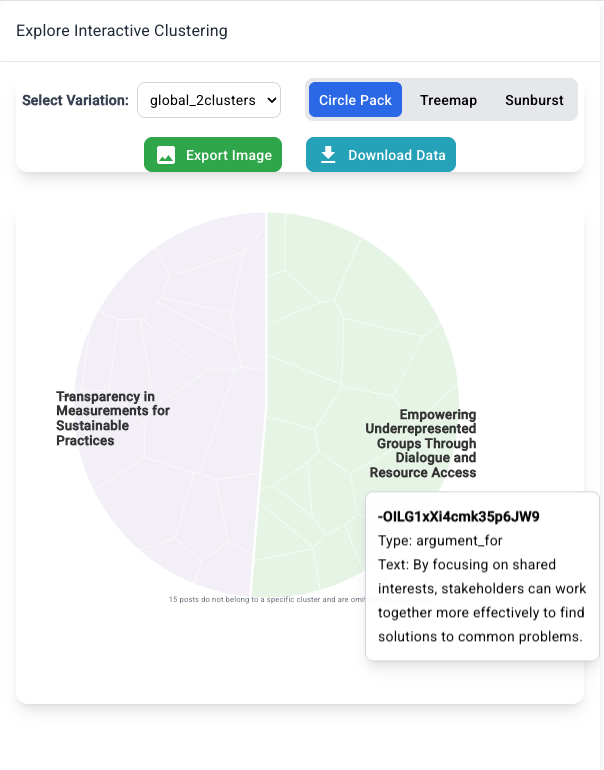}        
        \label{fig:subim1}
    \end{subfigure}%
    \hfill%
    \begin{subfigure}{0.3\textwidth}
        \centering
        \caption{4-Clusters}
        \includegraphics[width=\textwidth]{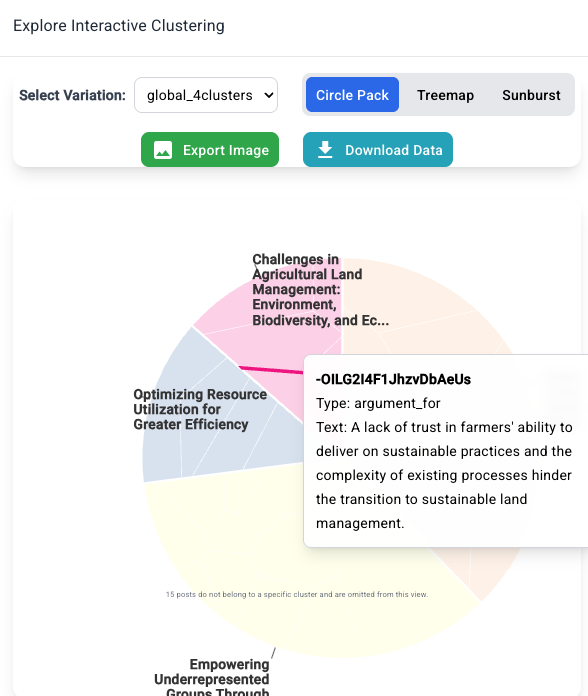}        
        \label{fig:subim2}
    \end{subfigure}%
    \hfill%
    \begin{subfigure}{0.3\textwidth}
        \centering
        \caption{8-Clusters}
        \includegraphics[width=\textwidth]{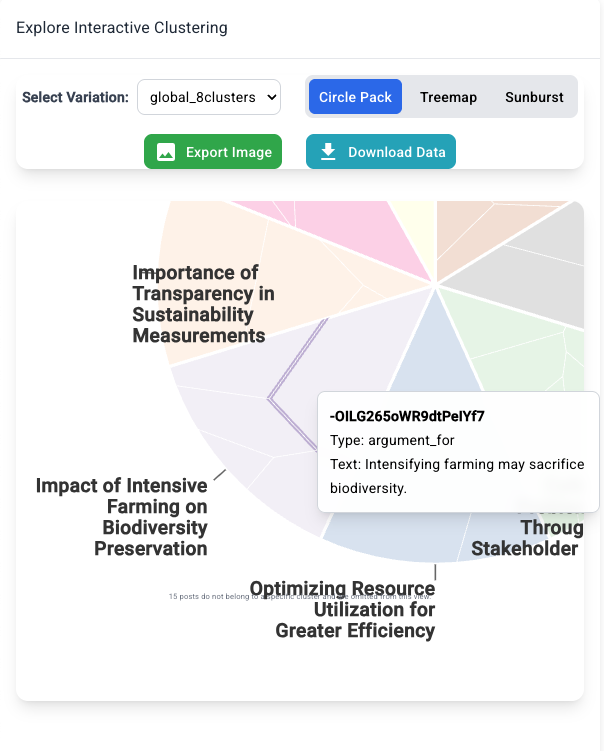}        
        \label{fig:subim3}
    \end{subfigure}    
    \label{fig:three_images}
\end{figure}

Advanced clustering algorithms group related arguments while preserving individual contributions, preventing the homogenization that often undermines True Representation in scaled deliberative processes. The interactive clustering visualisations, show in Figure~\ref{fig:three_images}, are powered by the same Fuzzy C-means algorithm used for policy recommendations. The user can interact with the visualisation, for example, by changing the number of clusters (from two to eight) and explore the debate in different levels of granularity, from broad themes to more nuanced sub-topics. The interactive clustering functionality allows users to navigate the discussion organised in a variable number of clusters, visualised either in interactive Voronoi, Treemap or Sunburst diagram.

\subsubsection{Theme mapping}
The 2D theme mapping functionality creates semantic similarity visualizations that allow users to navigate the deliberative space according to conceptual relationships (instead of chronological or hierarchical structures). This is achieved by applying a dimensionality reduction technique (UMAP) to the sentence embeddings of the arguments, plotting them in a 2D space where proximity relfects semantic similarity. This visualisation makes the hidden conceptual structure of the conversation explicit. The semantic mapping directly addresses the Collective Understanding Point of Struggle by revealing hidden connections between different perspectives, while supporting True Representation by ensuring that minority or outlier positions remain visible even when they don't cluster with majority views. 

These interconnected visualization capabilities collectively transform BCause from a linear discussion platform into a multidimensional deliberative space where collective intelligence emerges through transparent, user-controlled analytical processes. The system maintains democratic legitimacy as it preserves human provenance and ensuring AI augmentation rather replacement of human judgement, while enabling large scale deliberation.


\subsection{System Architecture and Integration}

\begin{figure}
    \centering
    \caption{BCause system architecture showing integration of data input sources (face-to-face transcripts, online contributions), AI processing components (Argumentation Mining, clustering, policy recommendation generation), and output interfaces (discussion platform, analytics dashboards, policy reports).}
    \includegraphics[width=0.95\linewidth]{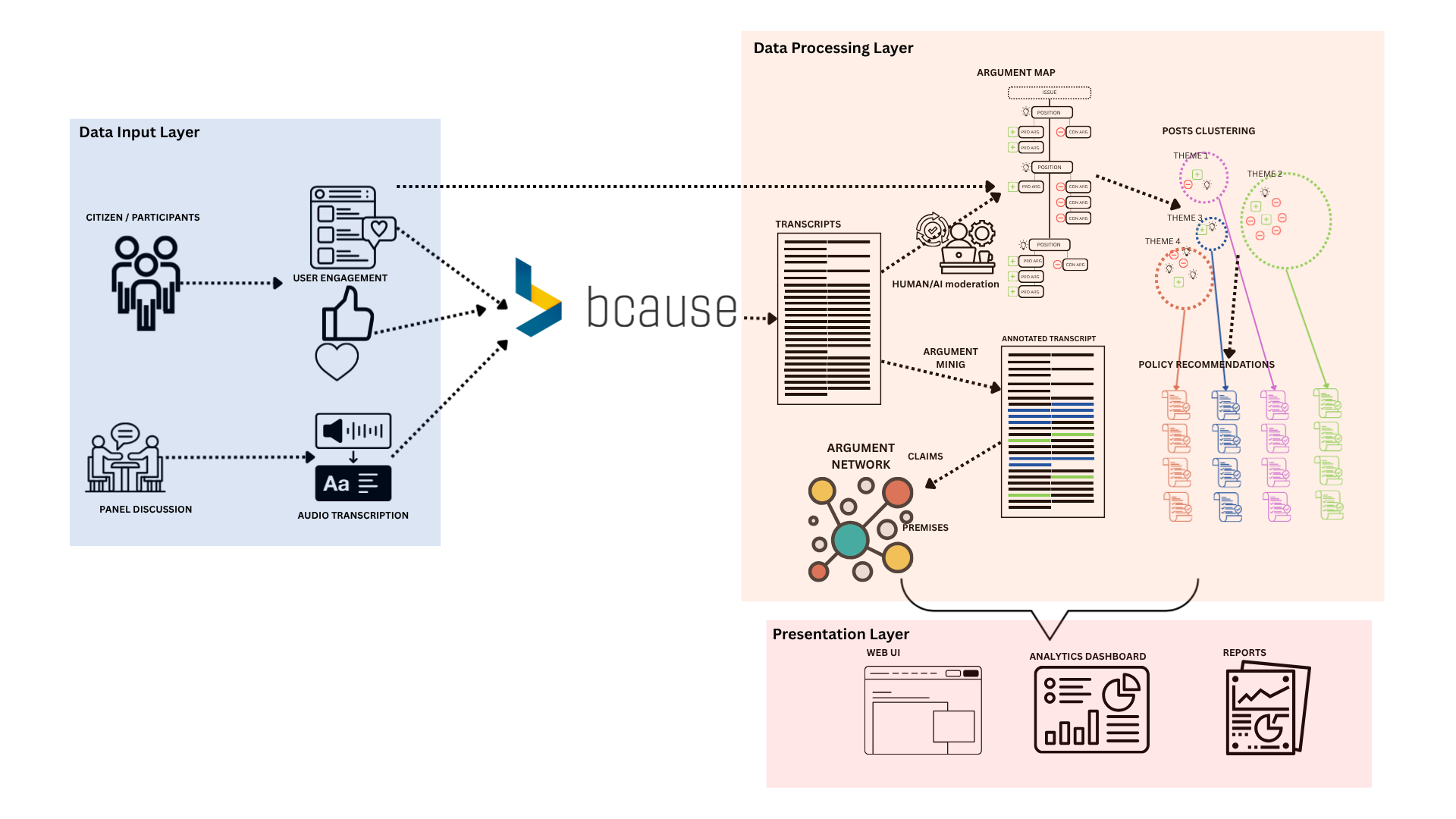}    
    \label{fig:bcause_architecture}
\end{figure}

Figure~\ref{fig:bcause_architecture} illustrates the BCause system architecture and its integration with external tools and technologies in the ORBIS use cases. The platform consists of three main layers: (1) the \textit{Data Input Layer}, which processes multiple sources including manual user contributions, automated transcript imports via speech-to-text APIs, and data from citizen dialogue events; (2) the \textit{AI Processing Layer}, which includes the argumentation mining pipeline, clustering algorithms, and LLM-based summarization and question generation modules; and (3) the \textit{Presentation Layer}, which delivers outputs through the web-based discussion interface, interactive analytics dashboards, and policy recommendation reports.

The integration workflow operates as follows: face-to-face deliberations from RIE Europe4Citizens events are recorded and transcribed using external speech-to-text services. These transcripts are then processed through the Argumentation Mining pipeline, which identifies IBIS components (Issues, Positions, Arguments). Human moderators review and approve the AI-generated structure before it is merged into the online BCause discussion space. Subsequently, the clustering algorithm groups semantically similar arguments, while the clusters themes generate policy recommendations displayed in the dashboard. Throughout this process, all transformations maintain provenance links back to original sources, ensuring transparency. 

\section{Hybrid Deliberation via DemocraticReflection: Enhancing Live Events Conversations with Real-Time Audience Engagement and Discourse Analysis}
\label{sec:dem_reflection}

While BCause addressed asynchronous deliberation challenges, DemocraticReflection represents a different approach to AI-augmented deliberation, focusing on real-time audience engagement during live or replayed events \cite{DeLiddo2020, CivicTechPeace2020} . 
This is quite relevant to use cases that engage citizens in live consultations sessions such as the CEPS Young Thinkers Initiative events. These live events exemplified a critical gap in our Points of Struggle analysis: how to maintain Collective Understanding and Transparency when deliberation occurs simultaneously across multiple channels—face-to-face expert presentations, live audience reactions, and digital participant engagement.

The Young Thinkers initiative traditionally relied on in-person consultations between young Europeans and EU policymakers, limiting participation to those who could physically attend Brussels meetings. This time, the initiative wanted to scale beyond its traditional expert consultation format to include larger numbers of young Europeans in policy deliberations with EU decision-makers. The challenge was not merely to scale participation, but to create meaningful integration between expert discourse and citizen input while preserving the dynamic, responsive nature of live deliberation. This required a tool that could simultaneously addresses all four Points of Struggle simultaneously: ensuring True Representation across diverse participation modes, fostering Collective Understanding between experts and citizens, maintaining Clarity and Transparency in how real-time input influences expert discussions, and achieving Integration and Scalability without losing the immediacy that makes live events valuable.

This drove the development and enhancement of DemocraticReflection as a ``second screen'' technology capable of capturing, analyzing, and synthesizing real-time audience engagement while providing immediate feedback to both facilitators and participants.

\subsection{DemocraticReflection: Platform Description}

\begin{figure}
    \centering
    \caption{Participant using DemocraticReflection to share their thoughts during a televised leadership election debate. See 3 min introductory video: https://youtu.be/xj0gB07yMuU}
    \includegraphics[width=1\linewidth]{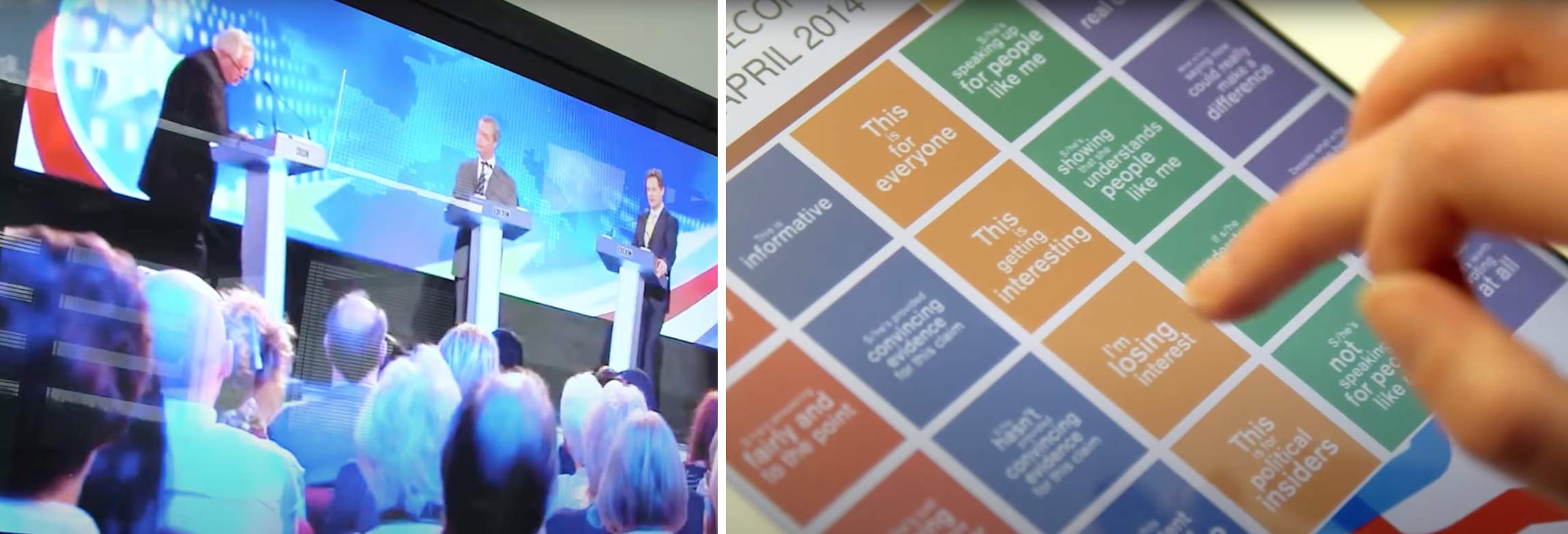}    
    \label{DR-ElectionDebate}
\end{figure}

DemocraticReflection\footnote{https://democraticreflection.cloud/} is a real-time interaction technology to crowdsource the Collective Intelligence of viewers making sense of live events or video replays~\cite{DeLiddo2020}. While watching either a live or replayed event, the audience interacts via a mobile phone or tablet by clicking on \textit{reflection cards} which signal the audience's feelings and immediate reactions in a dynamic way. The cards thus go far beyond the simplistic "thumbs up/down" sometimes used to gauge audience sentiment in live debates, and can be customised to the event, audience, and focal interest of the desired analytics. This trace of participants' experience is aggregated and analysed to provide insights into how the event was seen through the eyes of the audience. Figure \ref{DR-ElectionDebate} illustrates its use during deployment in the UK's 2010 election debate broadcasts. It has been successfully used to identify and challenge personal and collective biases \cite{Anastasiou2023}.

\subsection{ORBIS Use Case Implementation and Extensions}

\subsubsection{Audio Transcription and Content Analysis.}

The most fundamental extension developed for the CEPS Young Thinkers use case was comprehensive audio transcription with real-time content analysis. Live expert presentations are automatically transcribed and analyzed for thematic content, addressing  the Collective Understanding Point of Struggle by ensuring that participant responses can be contextualized within the expert discussion they reference.
The AI pipeline works as follows: a speech-to-text service (with speaker identification) transcribes the event audio and the resulting text is immediately fed to a fine-tuned LLM for thematic analysis. The latter part of the pipeline extracts the main themes of the discussion and summarises each speakers key positions. Then in combination with the audience engagement data (timestamped clicks on reflection cards), it generates a set of questions (open, clarifying, provocative) for each identified theme and another set of questions targeting each speaker. 

\begin{figure}
    \centering
    \caption{Customised Reflection Cards for CEPS Young Thinkers}
    \includegraphics[width=0.7\linewidth]{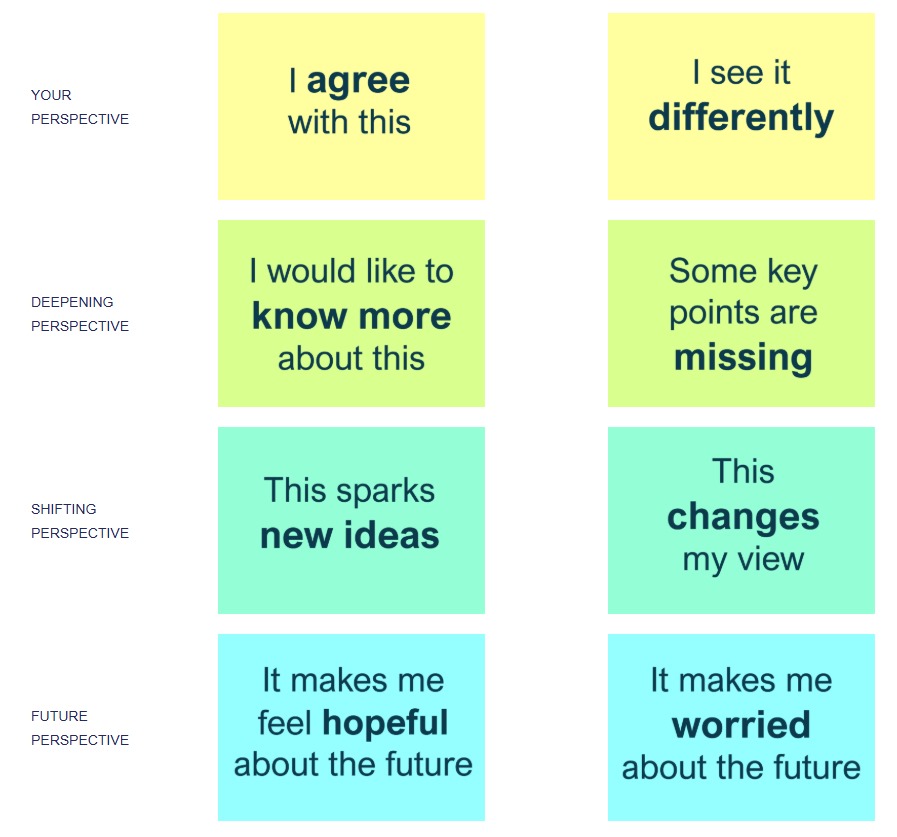}    
    \label{DR-RIE}
\end{figure}

These new features were tested during the CEPS flagship Ideas Lab event\footnote{https://www.ceps.eu/ceps-ideas-lab/} where they convened a live panel discussion of four experts while attended by a 25 people audience, on the topic of \textit{``AI for all: how to better design and regulate AI for fairness''}. The real-time analysis revealed critical moments where audience sentiment diverged from expert optimism. Notably, when the AI Liability Directive was discussed, the system detected a significant spike from the audience responses coded as 'It will not work' -- indicating immediate scepticism about regulatory effectiveness that might not have been apparent to facilitators focused on chairing the expert presentations.

The transcription system goes beyond simple speech-to-text conversion, employing NLP (Natural Language Processing) to identify key concepts and themes within experts presentations. This aligns with the Discussion Analysis and Visualization (DAV) system requirement -- participants can view not only what experts are saying, but how their contributions relate to the emerging thematic patterns of the discussion.

\subsubsection{Integrated Reflection and Content Synthesis}

\begin{figure}[ht]
    \centering
    \caption{The DemocraticReflection dashboard for the CEPS event showing two distinct interfaces: (a) a public view with detailed analytical visualizations and (b) a mid-event facilitator view with discussion summaries and engagement metrics.}
    \begin{subfigure}{0.49\textwidth}
    \centering
    \includegraphics[width=0.99\textwidth]{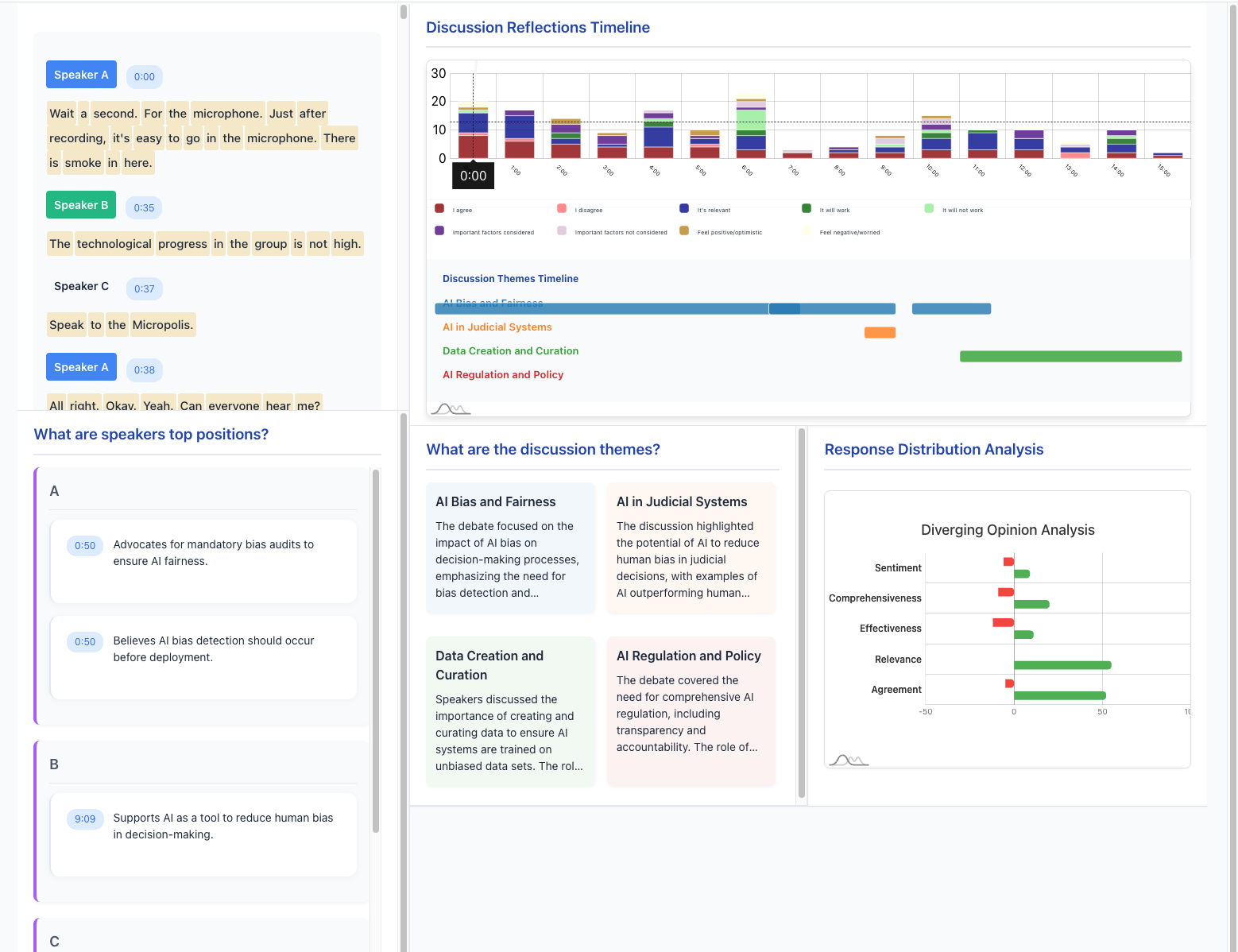}
\end{subfigure}
\hfill
\begin{subfigure}{0.49\textwidth}
    \centering
    \includegraphics[width=0.99\textwidth]{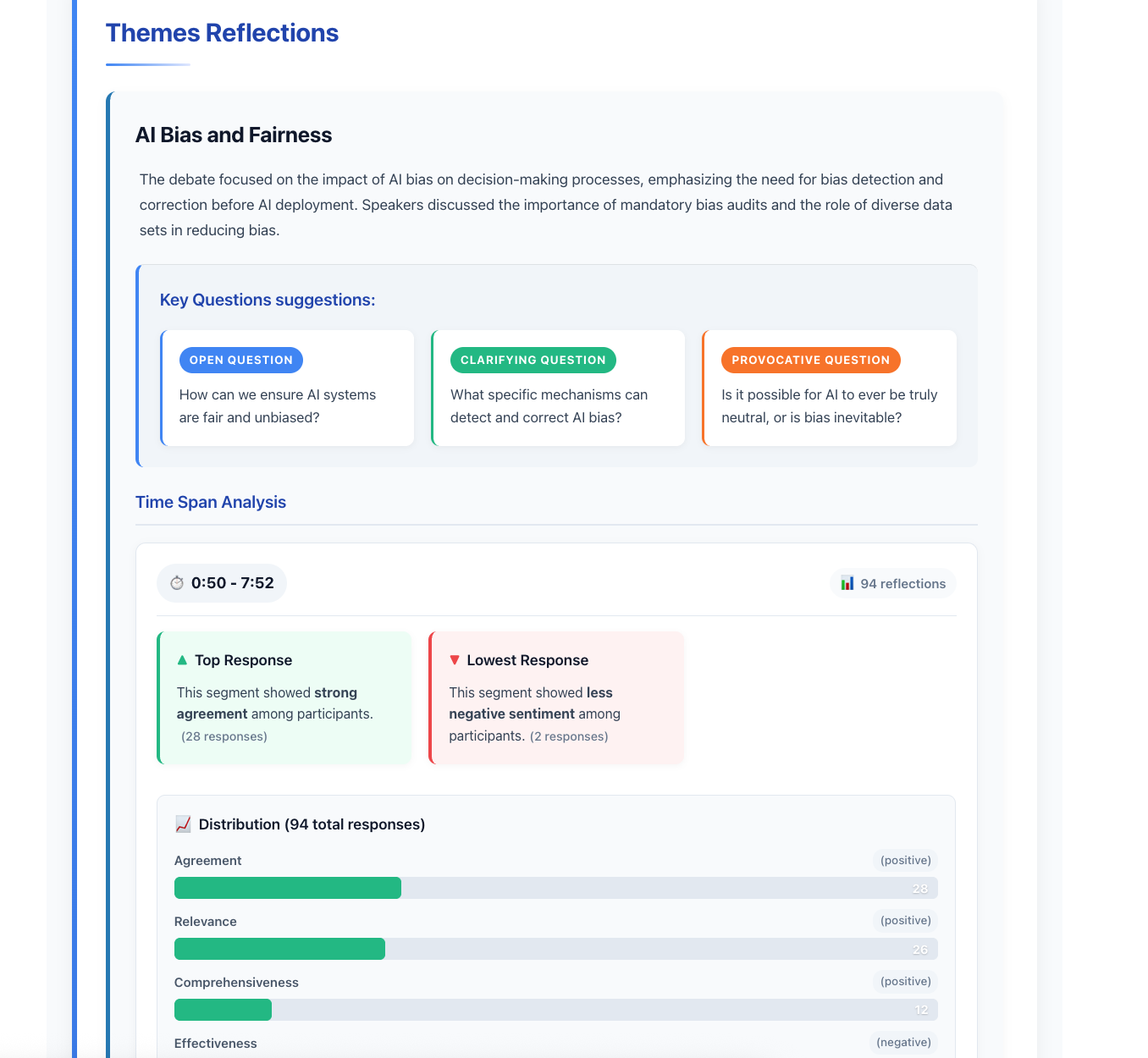}
\end{subfigure}
\label{fig:demo-reflection-public}
\end{figure}

These aggregated audience reflections are combined with transcript content to create dynamic textual summaries plus a thematic timeline, illustrated in the Democratic Reflection dashboard (Figure~\ref{fig:demo-reflection-public}). This data fusion addresses the True Representation Point of Struggle, by taking into account each citizen's input and ensuring that it stays connected with experts' opinions, preventing marginalising public input in policy discussions.
Each participant interaction is timestamped and linked to specific moments in the live panel discussion, expert presentation or any other physical face-to-face live event. This rich dataset shows not just \textit{what} citizens think, but \textit{when and why} they formed those opinions. This fulfills the ``Report and Summarization'' and ``Clarity and Transparency'' system requirements, by making the relation of policy-making discourse and citizen feedback explicit and traceable.

\subsubsection{Dynamic Question Generation for Dialogue Continuity }

DemocraticReflection provides two separate views in a live event: the public \textit{audience }view and the private \textit{facilitator} view for the moderators/organisers. The facilitator view (Figure \ref{fig:demo-reflection-public}b) is where a key generative AI feature comes into play: \textit{dynamic question generation}. The AI system identifies moments of significant divergence or consensus between the expert discussion and audience reactions. It then uses a Generative LLM to draft questions designed to bridge these gaps. The AI is prompted with specific data: a transcript segment, the associated audience reflection data, and an instruction (e.g., ``Based on the speaker's point and the audience's strong 'disagreement' reflection, generate a clarifying question for the facilitator to ask''). This provides human moderators, who are already under high cognitive load managing a live event, with data-driven, contextually relevant prompts to foster a more inclusive and responsive dialogue. This directly supports human moderators rather than replacing them, fulfilling the "Moderation and Assistance" (MA) requirement. For instance, during the CEPS Ideas Lab, the system identified tension between an expert's discussion of community-generated data and audience concerns about ethics. It generated the question:\textit{ "Is it ethical to rely on community-generated data for AI systems?"} which the facilitator then posed to the panel, bringing the aggregated audience sentiment into the expert dialogue.

To summarise, the above extensions collectively transformed DemocraticReflection from an aggregated feedback tool into a more comprehensive platform for hybrid CI, bridging the temporal and contextual gaps between experts and citizens -- critically,  without disrupting the tempo and sense of audience connection which are such essential features for meaningful democratic consultation in hybrid settings.

\section{Conclusions}
\label{sec:conclusions}

In this chapter, we have described a comprehensive human-centred design process for eliciting user requirements for Deliberative Democracy technologies, with particular attention to responsible use of AI. This approach gives a meaningful voice to stakeholders in shaping requirements. We framed this conceptually in terms of \textit{collective intelligence} for deliberative democracy, bringing human and machine intelligence into dialogue. 

In close partnership with organisations and communities working on advocacy and citizen-led policy making, we surfaced their nuanced challenges and aspirations for future Deliberative Democracy systems. Four pivotal ``Points of Struggle'' emerged, which essentially represent the four core problems that DD approaches should aspire to solve, serving as the drivers for defining new scalable AI-augmented deliberation platforms. This co-creative approach motivated user scenarios and high level system requirements intended to meet diverse aspirations, and address the sociotechnical challenges faced by both technical and non-technical stakeholders. The should guide the design and implementation of future CI tools for DD. We have therefore documented how the Points of Struggle and system requirements were translated in the design of BCause and DemocraticReflection, in the context of authentic deployments with DD organisations.

As introduced at the start of this chapter, throughout this research program, we have sought to answer the question of how we can tell if a system is exhibiting CI, adapted from Gupta et al.~\cite{Gupta2023}: \textit{``How do we know that a DD sociotechnical system as a whole, consisting of a complex web of hundreds of human–machine interactions, is exhibiting democratic results, impacts and behaviours?''.} Our two exemplars, BCause and DemocraticReflection, are not merely technology demonstrations; they are instruments for observing CI4DD in action. With BCause, we see evidence of collective memory and reasoning. We know CI4DD is being exhibited when we can trace the journey of an idea from a face-to-face conversation transcript into a structured online argument, see it clustered with semantically similar ideas from other users (DAV.1), and watch it become part of a synthesised policy recommendation (RS.2). With DemocraticReflection, we see evidence of collective attention and real-time sensemaking. We know CI4DD is present when the system detects a critical divergence between an expert's statement and the audience's immediate, unspoken sentiment (RS.6), and then generates a clarifying question (MA.5) that a human facilitator uses to steer the conversation. We are witnessing CI4DD in the system's ability to create a transparent, persistent, and structured map of the group's thinking that is greater than the sum of its individual posts, and in the system’s capacity to create a feedback loop that makes the ``sense of the room'' visible and actionable, shaping the dynamics of the live deliberation event.

Returning to our initial conceptualisation of CI4DD, and in particular our focus on AI-augmentation of DD, we note Clark's (2025) recent contextualisation to generative AI of his influential work on ``extended mind''. Clark argues that the ability to engage with LLM-based dialogic agents is simply another such extension in the history of humanity's use of tools. The crux of the matter is that such extensions to our minds must be designed and used judiciously: 

\begin{displayquote}
\textit{``The lesson is that it is the detailed shape of each specific human-AI coalition or interaction that matters. The social and technological factors that determine better or worse outcomes in this regard are not yet fully understood, and should be a major focus of new work in the field of human-AI interaction.''}
\end{displayquote}

Our hope is that this chapter clarifies what this can look like in the context of hybrid CI4DD. In the CI4DD framework (Figure~\ref{CI4DD}), AI is integrated as one of many actors in a complex human-machine collaboration. CI4DD can emerge through human-AI interaction when technologies and processes are co-conceived, co-designed, and orchestrated coherently via human-centered design. 

New theories, technical advances, and future evaluation studies should further map the CI4DD research and design spaces. While AI is undeniably a disruptive force in society, we propose that it may also be harnessed to augment our collective intelligence in an increasingly hybrid human-AI world, which will continue to challenge our democratic processes.

\clearpage

\bibliographystyle{apalike}
\bibliography{sample}

\section*{Acknowledgements}
This research was funded in collaboration by UKRI under the UK Government’s Horizon Europe Guarantee scheme (Reference Number: 10048874) and by the European Commission under the Horizon Europe Programme, in the context of the ORBIS Project (GA: 101094765) on ``Augmenting participation, co-creation, trust and transparency in Deliberative Democracy at all scales''.

\section*{Appendix: High-Level System Requirements}

This appendix provides the full catalogue of 36 high-level system requirements derived from the co-design process. They are organized into the six functional categories referenced in the main text.

\paragraph{1. User Interaction and Engagement (UIE)}
Requirements that involve the overall user experience and engagement with the system.
\begin{itemize}
\item UIE.1: The system stores user profiles.
\item UIE.2: Notification mechanism for event updates.
\item UIE.3: Personal space (idea resource pad) for each user.
\item UIE.4: User-friendly interface for exploring network graphs of ideas.
\item UIE.5: Real-time transcription of live discussion.
\item UIE.6: User recommendation according to user profile.
\end{itemize}

\paragraph{2. Discussion Analysis and Visualization (DAV)}
Requirements regarding the examination and representation of textual discussions.
\begin{itemize}
\item DAV.1: Cluster discussion data into main viewpoints/arguments.
\item DAV.2: Summarization of main arguments of discussion.
\item DAV.3: Identification of key influential people/actors.
\item DAV.4: Comparison of argument graphs and detection of overlaps and disjoints.
\item DAV.5: Real-time analysis and classification of key elements of discussion.
\item DAV.6: User-friendly interactive interface for exploring viewpoints.
\end{itemize}

\paragraph{3. Moderation and Assistance (MA)}
Requirements focusing on the facilitation of effective collaborative discussion.
\begin{itemize}
\item MA.1: Expert collaboration space to assist policy formulation.
\item MA.2: Minutes creation and structured discussion interface.
\item MA.3: Feedback mechanisms (e.g., voting, thumbs up/down, etc.).
\item MA.4: Controversy detection.
\item MA.5: Automated indicators and key points to assist moderators.
\item MA.6: Event evolution infographic generator.
\end{itemize}

\paragraph{4. Reports and Summarization (RS)}
Requirements centered around the generation of informative summary reports.
\begin{itemize}
\item RS.1: Different styles/tone of generated reports.
\item RS.2: Abstractive summarization and statistical or analytical data.
\item RS.3: Generates reports on the implicit impact of discussions.
\item RS.4: Analytics reports and statistical analysis reports.
\item RS.5: Summary report generator that maintains source links.
\item RS.6: Real-time audience feedback analysis.
\end{itemize}

\paragraph{5. Collaborative Features (CF)}
Requirements fostering collective editing and collaboration among users.
\begin{itemize}
\item CF.1: Idea filtering and grouping.
\item CF.2: Mini-voting mechanism (e.g. majority voting).
\item CF.3: Connection and query external knowledge bases.
\item CF.4: Merge and prioritization of key-points.
\item CF.5: Recommendation of experts according to the topic of deliberation.
\item CF.6: Social media sharing.
\end{itemize}

\paragraph{6. Multi-Phase Deliberations (MPD)}
Requirements for supporting deliberation processes carried out in multiple stages.
\begin{itemize}
\item MPD.1: Support various types of multi-phase deliberations.
\item MPD.2: Polling mechanism for advancing deliberation stages.
\item MPD.3: Aggregate survey results.
\item MPD.4: Historical record of deliberation and aggregation of past deliberations.
\item MPD.5: Issue-to-proposal generator and composer of convincing policy change pitch.
\item MPD.6: Prediction of group acceptance and projection into an embedded space of key attributes.
\end{itemize}

\end{document}